\documentclass{PoS}

\usepackage{multirow, booktabs}
\usepackage{siunitx}
\usepackage{url}
\usepackage{graphicx}

\usepackage{amssymb}
\usepackage{fontenc}
\usepackage{times}
\usepackage{mathptmx}
\usepackage{graphicx}

\graphicspath{ {/./}}
\DeclareGraphicsExtensions{.png,.pdf}

\newcommand{\fermi}{$Fermi$}
\newcommand{\swift}{$Swift$}
\newcommand{\tburst}{$t_{\mathrm{Burst}}$}
\newcommand{\g}{$\gamma$}

\title{GRB Observations with H.E.S.S. II}

\ShortTitle{GRBs with H.E.S.S. II}

\author{\speaker{Clemens Hoischen} $^{1}$,
Arnim Balzer $^{2}$,
Elisabetta Bissaldi $^{3}$,
Matthias Fü{\ss}ling $^{4}$,
Tania Garrigoux $^{5}$,
Daniel Gottschall $^{6}$,
Markus Holler $^{7}$,
Alison Mitchell $^{8}$,
Paul O'Brien $^{9}$,
Robert Parsons $^{8}$,
Gerd Pühlhofer $^{6}$,
Gavin Rowell $^{10}$,
Fabian Schüssler $^{11}$,
P.H. Thomas Tam $^{12}$,
Stefan Wagner $^{13}$,
for the H.E.S.S. Collaboration \\
$^{1}$ Institut f\"ur Physik und Astronomie, Universit\"at Potsdam,  Karl-Liebknecht-Strasse 24/25, D 14476 Potsdam, Germany \\
$^{2}$ GRAPPA, Institute of High-Energy Physics, University of Amsterdam,  Science Park 904, 1098 XH Amsterdam, The Netherlands \\
$^{3}$ Instituto Nazionale di Fisica Nucleare  \\
$^{4}$ DESY, D-15738 Zeuthen, Germany\\
$^{5}$ Centre for Space Research, North-West University, Potchefstroom 2520, South Africa\\
$^{6}$ Institut f\"ur Astronomie und Astrophysik, Universit\"at T\"ubingen, Sand 1, D 72076 T\"ubingen, Germany  \\
$^{7}$ Institut f\"ur Astro- und Teilchenphysik, Leopold-Franzens-Universit\"at Innsbruck, A-6020 Innsbruck, Austria \\
$^{8}$ Max-Planck-Institut f\"ur Kernphysik, P.O. Box 103980, D 69029 Heidelberg, Germany \\
$^{9}$ Department of Physics and Astronomy, University of Leicester, University Road, Leicester, LE1 7RH, United Kingdom.  \\
$^{10}$ School of Physical Sciences, University of Adelaide, Adelaide 5005, Australia \\
$^{11}$ IRFU, CEA, Universit\'e Paris-Saclay, F-91191 Gif-sur-Yvette, France  \\
$^{12}$ Institute of Astronomy and Space Science, Sun Yat-Sen University, Guangzhou 510275, China  \\
$^{13}$ Landessternwarte, Universit\"at Heidelberg, K\"onigstuhl, D 69117 Heidelberg, Germany \\
E-mail: \email{contact.hess@hess-experiment.eu}}

\abstract{The High Energy Stereoscopic System (H.E.S.S.) has been searching for counterparts of Gamma Ray Bursts (GRBs) for many years. In 2012 the system was upgraded with a fifth 28\,m diameter telescope (CT5) which is equipped with faster motors for rapid repointing, marking the start of the second phase of H.E.S.S. operation (H.E.S.S. II). CT5s large light collection area of $600\,\mathrm{m}^{2}$ improves the sensitivity to low-energy gamma-rays and even extends the energy range below 100 GeV. The search for counterparts continues now in the energy range of tens of GeV to tens of TeV. A detection in this energy range would open a new window to the part of the spectrum of these highly energetic explosions which \fermi-LAT has only successfully detected in a reduced subset of events, with rather limited statistics. In the past years, H.E.S.S. has performed follow-up observations based on GRB detections by \swift-BAT and \fermi-GBM/-LAT. This Target of Opportunity observation program was carried out with a generalised Target of Opportunity Alert system. This contribution will highlight key features of the Target of Opportunity Alert system, present follow-up statistics of GRBs as well as detailed results of promising follow-up observations.}

\FullConference{35th International Cosmic Ray Conference --- ICRC2017\\
		10-20 July, 2017\\
		Bexco, Busan, Korea}

\begin{document}

\section{Introduction \& Motivation}
Gamma Ray Bursts (GRBs) are among the most violent phenomena known in the universe emitting up to $\rm 10^{55} erg$ within typically only a few seconds. They are seen as short intense periods of \g-ray emission in the keV to MeV energy range lasting from as little as $0.1$\,s up to a few hundred seconds. This intense emission phase is called the {\it prompt} phase of the GRB and is followed by a so called {\it afterglow} phase during which the emission decays on a longer timescale. During the afterglow phase of the GRB, counterparts in all wavelengths are commonly detected, from radio to optical. An explanation of these phases is provided in the so called {\it Fireball-Model}\,\cite[and references therein]{fireball}. In this model, particle acceleration takes place at multiple points of the systems evolution. It assumes that the formation of a new compact object produces an ultra-relativistic jet that is powered by the compact object as the central engine. The engine emits plasma shells with different Lorentz factors $\Gamma(\rm t)$ which will eventually collide. During every collision of two shells, shock acceleration, giving rise to \g-rays in a broad energy range from keV- potentially up to TeV energies can be produced. At some point, the shells (and the jet) will hit the unshocked interstellar medium. This collision is thought to be the start of the afterglow phase during which shock acceleration can take place again. If the \g-radiation produced in these various stages of the phenomenon can actually escape from the object heavily depends on the photon field density which, in the case it is too high, will lead to \g-\g~absorption that effectively cascades down the initial energy of the photons. Similarly, the highest energy photons can be absorbed through interactions with the Extragalactic Background Light (EBL) if the distance between the object and the observer is too large. \\

Nowadays, GRBs with a confirmed High-Energy (HE, $\rm E_{\gamma} > 100$\, MeV) \g-ray emission are rare. A confirmed component above energies of $\rm 100\, MeV$ is present in only 5\,\% of the detected GRBs, while above energies of $\rm 10\, GeV$ this fraction is further reduced to below 1\,\%\cite{lat_he_grbs}. The GRBs with a detected high energy component are the ones that challenge the current {\it fireball-model} paradigm the most, demanding extreme gamma factors and high acceleration efficiencies. Thanks to the thousands times larger effective area, ground based \g-ray astronomy allows for counterpart searches with several orders of magnitudes better sensitivity on short timescales (below a few hours) at energies around $\rm 100\, GeV$\,\cite{cta_sens_time}. A detection of a GRB at these energies would therefore further challenge the current understanding of the GRB phenomenon and would provide a deeper insight into particle acceleration in GRBs.\\

Even though the ground-based \g-ray community is trying for quite some time to find counterparts above energies of $100$\,GeV, this effort remains unsuccessful to date. H.E.S.S. reported about the observations taken in the GRB follow-up program in the years 2003 to 2007\,\cite{hess-grb-2003-2007}. The same program is still running and will likely be continued until the decommissioning of H.E.S.S. Several improvements to the system were made in the past: In 2012 a fifth Telescope was deployed: CT5 has the largest light collection area in ground-based gamma-ray astronomy with $600\,\mathrm{m}^{2}$ which lowers the energy-threshold down to around $50$\,GeV for good observation conditions. At these energies the extragalactic background light (EBL) does not attenuate the \g-rays significantly up to redshifts of around $z=1.5$\,\cite{ebl_gilmore}. Additional technical improvements in the alert system and the follow-up procedures will be described in the following section.\\

\section{Follow-up Procedures}
GRB observations by H.E.S.S. are issued by a fully automated system that is integrated into the central control software\,\cite{hessdaq} of the experiment. The automation ranges from the reception of alerts to preliminary analysis results. This system is a software chain which receives alerts\footnote{We make use of the tool {\it comet}\,\cite{comet}} in the {\it VOEvent}\,\cite{seaman2011ivoa} format defined by the {\it International Virtual Observatory Alliance\footnote{see \url{www.ivoa.net}}}. All alerts pass through a processing and filtering facility with each alert being correlated to science requirements defined by the Collaboration. Only alerts that fulfil all specified follow-up requirements will be forwarded to the central telescope steering software which is able to automatically override the current observation schedule in order to observe the burst location. The data that is taken is analysed automatically by a preliminary analysis in real time. Based on these preliminary results the prolongation of the observations are being decided. The system is not restricted to GRB follow-up observation but also enables us to promply follow-up on all kinds of Multi-Messenger transient targets (see \pos{PoS(ICRC2017)653}).\\

Time-critical observations such as GRB follow-up observations will make use of a special repointing procedure which reduces the necessary time to restart the observations. Observations will start as soon as the target location is inside the field of view of CT5, regardless of the status of CT1-4 (which can join the observation afterwards)\,\cite{balzer1}. During the development of CT5, a fast repointing speed was one key requirement. In order to allow for even faster repointing speeds, CT5 is allowed to slew overhead through the zenith in the so called {\it "reverse"} mode, if it is the fastest way to reach the target position\,\cite{hofverberg2013}.\\

In the GRB follow-up program we perform observations based on GRB detection from \fermi-LAT\,\cite{fermilat}, \fermi-GBM\,\cite{fermigbm} and \swift-BAT\,\cite{swiftbat} which are all being provided by the Gamma-ray burst coordinate Network (GCN)\,\cite{gcn}. In detail we make use of the following alert types:
\begin{itemize}
\item \url{Fermi#GBM_Gnd_Pos},
\item \url{SWIFT#BAT_GRB_Pos},
\item \url{Fermi#LAT_Updated_Pos} and
\item \url{Fermi#LAT_Offline_Pos}.
\end{itemize}

Multiple sets of follow-up requirements are connected to each alert type. If the timescale on which the burst location becomes visible for H.E.S.S. is short enough, the observations will start automatically while making use of the accelerated transition scheme mentioned before which allows CT5 to slew in reverse mode. If the position becomes visible not immediately, but with less than four hours delay with respect to the burst time \tburst, the shifters will schedule the observations manually. For locations that become visible more than 4 hours after the burst, an expert on-call will be informed in order to decide if H.E.S.S. can add valuable information by observing this individual burst regardless of the long delay.\\

The full system is regularly tested using self issued alerts that enter the system in the same way the GRB alerts do. This allows for stable operations of all sub-components of the system.

\section{Follow-up Statistics}
\begin{figure}[htbp]
\begin{center}
\includegraphics[width=\textwidth]{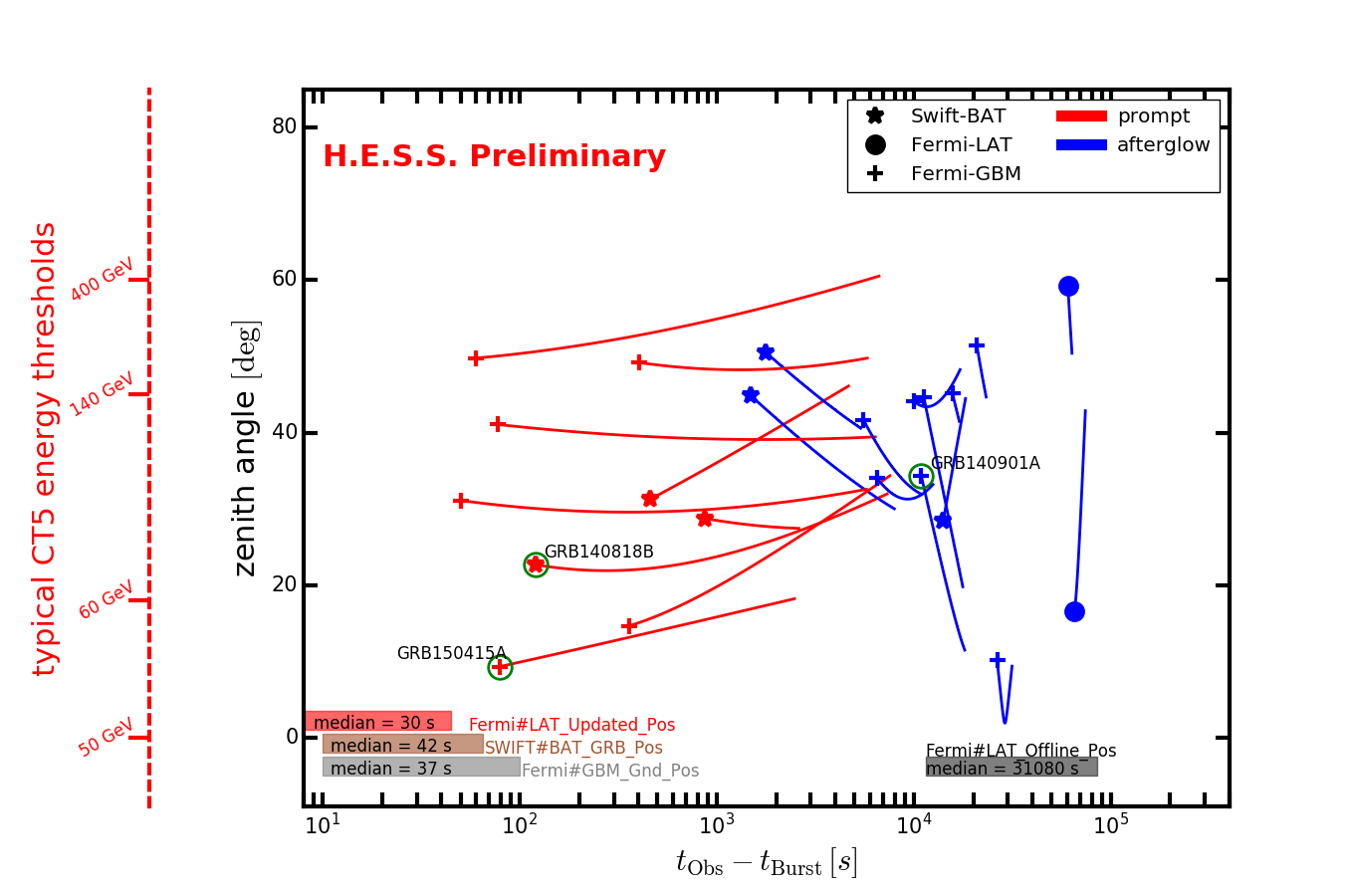}
\caption{Overview of H.E.S.S. GRB follow-up observations from 2012 to may 2017 for which CT5 was available. Each marker denotes the start of an observation. The line following the marker denotes the evolution of the zenith angle under which the final location of the burst was visible from the H.E.S.S. site at the time of the observation. The typical delay timescales for every alert type was evaluated using the public alert archive from 2015 to 2017 and is shown at the bottom of this figure. This delay is the baseline  for the the H.E.S.S. reaction and repointing speed. Red markers denote observations that were taken in a fully automated way, while the the blue ones were scheduled by hand. The additional red axis denotes the energy threshold that roughly corresponds to the respective zenith angle.}
\label{fig:follow-zen-delay}
\end{center}
\end{figure}

More than 30 GRB detections by \fermi\,and \swift\,were followed up by H.E.S.S. since 2012. CT5 was not available for all observations. Among the reasons are technical difficulties or unclear weather situations which make observations with CT5 not possible. The observations for which CT5 was available are displayed in Figure~\ref{fig:follow-zen-delay}. Each follow-up is shown in the plane of the zenith angle under which H.E.S.S. could observe the final burst position and the time of the observation with respect to \tburst. The delay between the observation start and \tburst is a sum of the time necessary for i) receiving and processing of the alert and repointing of the telescopes, ii) the detection of the event itself and the distribution of it which involves the downlink from the satellites to the ground  and ultimately to the H.E.S.S. site. The delay of each individual alert is the baseline for all H.E.S.S. follow-up opportunities. The typical delay ranges for the distribution of the different alert times utilised in the GRB program are highlighted on the bottom of Figure~\ref{fig:follow-zen-delay} and were derived based on the history of alerts from 2015 on. The green circles highlight the three bursts that will be discussed in more detail. Reaction times of less then 100 seconds are achieved on a regular basis.  

\section{Analysis Results of selected Observations}
\subsection{Selected Observations}
In the following, three follow-up observations of GRBs will be discussed in more detail:
\begin{itemize}
\item \textbf{GRB140818B}, a long burst detected by \swift-BAT, followed up by H.E.S.S. promptly with a delay of $120$\,s with respect to the burst time. The location of the burst was observed for two hours with an average zenith angle of $25$\,deg.
\item \textbf{GRB140901A}, a short burst detected by \fermi-GBM, followed up with a delay of roughly 3 hours. The positional uncertainty quoted for this burst of $2.5$\,deg is just as large as the H.E.S.S. II field of view. The position was also observed for two hours with an average zenith angle of $22$\,deg.
\item \textbf{GRB 150415A}, a mid duration burst ($\mathrm{T_{90}} = 8\,{\rm s}$) detected by \fermi-GBM, followed up promptly. Observations started only $80$\,s after the detection by \fermi-GBM. The position was updated which resulted in a repointing of the H.E.S.S. array after the first 28 minutes of observations. This final position was then observed for the remaining 17 minutes of the night. The position uncertainty of this burst was quite large with $3.6$\,deg and exceeded the size of the CT5 field of view. The location was observable for roughly 45 minutes with an average zenith angle of $14$\,deg.
\end{itemize}
Additional information on these bursts is given in table~\ref{tab:grbsummary}. Unfortunately, for neither of the three bursts a redshift could be measured. None of the three observations yielded a detection. Nevertheless they highlight nicely the potential, but also the intrinsic difficulties in the GRB follow-up program.

\begin{table}[t]
\begin{center}
\begin{tabular}{c||c|c|c}
\midrule[2pt]
						&	GRB140818B	& GRB140901A	& GRB150415A \\
\midrule[2pt]
\multicolumn{4}{c}{\bf General information on the GRB detection}	\\
Observations triggered by		&	\swift-BAT		& \fermi-GBM		& \fermi-GBM	\\
$\rm T_{90}[s]$				&	$18.1$		 & $0.16$			& $8.0$		\\
Also detected by			&	UVOT, XRT,IPN & IPN			& $-$		\\
Significance				&	$10\,\sigma$	 & $68\,\sigma$		& $19\,\sigma$ \\
Redshift					&	{\bf $ \times$}	 & {\bf $ \times$}	& {\bf $ \times$}\\
\hline
\multicolumn{4}{c}{\bf Final position of the GRB}	\\
Right-Ascension (J2000) [deg]	&	$217.15$		& $15.82$			& $220.63$	\\
Declination (J2000) [deg]		&	$-1.36$		& $-32.76$		& $-19.34$	\\
Position uncertainty [deg]		&	$0.05$		& $2.5$ 			& $3.6$		\\
\hline
\multicolumn{4}{c}{\bf H.E.S.S. Observation parameters} \\
Average zenith angle [deg]		  	&	25			& 22				& 13.7	\\
Delay of observation start w.r.t \tburst 	&	120 s		& 3 hours			& 80 s	\\
Observation time				  	&	2 hours		& 2 hours			& 40 minutes \\
Fully automated reaction				&	\checkmark	& {\bf $ \times$}	& \checkmark \\
\midrule[2pt]
\end{tabular}
\end{center}
\caption{Summary of information about the three selected H.E.S.S. GRB follow-up observations and their detection in X-rays.}
\label{tab:grbsummary}
\end{table}%

\subsection{Results}
The observations for all three bursts did not reveal any significant emission in the respective fields of view. The observation triggered by \swift-BAT (GRB140818B) has an accurate enough localisation so that the differential flux upper limits could be calculated at the location of the \swift-BAT detection. For GRB140901A and GRB150415A, detected by \fermi-GBM, the H.E.S.S. fields of view only cover 68\,\% of the GBM localisation at best. Therefore the limits obtained are given at the best fit location as well as $1.5$\,deg off the center of the field of view in the case of GRB140901A. In the case of GRB150415A, the limit at the best fit location could only be calculated for the second part of the observation due to the coordinate update. Therefore we present the differential flux upper limits at the intersection of both observed fields, as well as at the best fit position of the burst. The limits at these different locations can be seen as a benchmark sensitivity that H.E.S.S. obtains for observations in similar observing conditions.

\begin{figure}[htbp]
\begin{center}
\includegraphics[width=\textwidth]{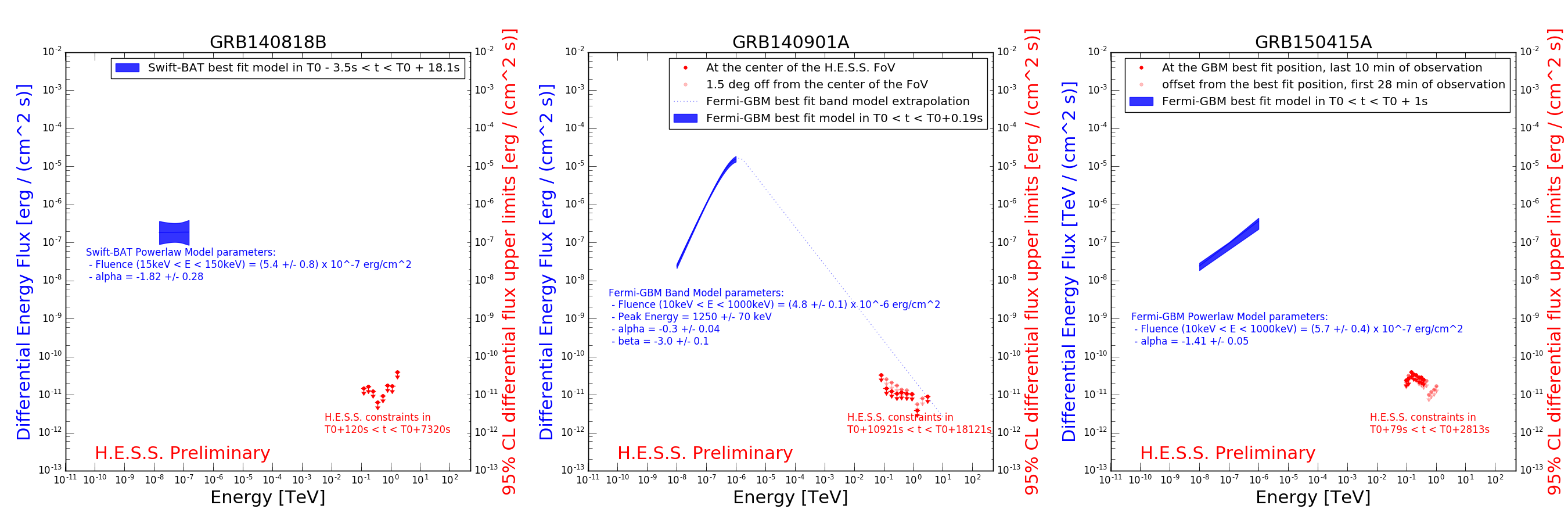}
\caption{The upper limit results obtained from the H.E.S.S. observations for all three bursts (GRB140818B, GRB140901A and GRB150415A from left to right) are shown in red. The best fit spectral model for the detection of each burst is shown in blue. For the two observations triggered by \fermi-GBM, upper limits obtained off of the center of the field of view are shown as well.}
\label{fig:grbuls}
\end{center}
\end{figure}

The observations were analysed using the {\it model-analysis}\,\cite{modelanalysis} and were cross-checked with an independent calibration, reconstruction and analysis chain\,\cite{impact}. The differential flux limit calculation was done by assuming that the potential \g-ray emission follows a power-law in energy with a spectral index of 2. Due to the low statistics of the observations, the limits were computed by rescaling the photon counts off of the target region to the live-time in the target region. This avoids random fluctuations and the resulting differential limits can be understood as the sensitivity of H.E.S.S. II for such observations. The limits are presented in Fig.~\ref{fig:grbuls}.

The differential flux upper-limits are on the level of $10^{-11}\,{\rm erg\,cm^{-2}\,s^{-1}}$, depending on the energy. For all three bursts an energy threshold of less then 100 GeV was achieved.

\section{Conclusions \& Outlook}
H.E.S.S. in its second phase will continue to perform follow-up observations of GRBs detected in the keV to MeV energy range by \fermi \,and \swift. By employing a fully automated system H.E.S.S. is able to react as fast as possible to GRB detections. CT5 will remain the world largest Cherenkov Telescope, possessing great discovery potential for GRBs due to the low energy threshold and its rapid slewing speed. Since there is no gap in energy to the \fermi-LAT experiment, the first detection of a GRB using the Imaging Air Cherenkov Telescope method is a matter of luck and time. Due to the large effective area of the Imaging Air Cherenkov Telescope method, H.E.S.S. might also be able to detect GRBs that are too faint for \fermi-LAT on the typical GRB timescales of at most a few hundred seconds. The follow-up procedures and requirements are being optimised to target specific physics questions.

\section*{Acknowledgements}
The support of the Namibian authorities and of the University of Namibia in facilitating the construction and operation of H.E.S.S. is gratefully acknowledged, as is the support by the German Ministry for Education and Research (BMBF), the Max Planck Society, the German Research Foundation (DFG), the Alexander von Humboldt Foundation, the Deutsche Forschungsgemeinschaft, the French Ministry for Research, the CNRS-IN2P3 and the Astroparticle Interdisciplinary Programme of the CNRS, the U.K. Science and Technology Facilities Council (STFC), the IPNP of the Charles University, the Czech Science Foundation, the Polish National Science Centre, the South African Department of Science and Technology and National Research Foundation, the University of Namibia, the National Commission on Research, Science \& Technology of Namibia (NCRST), the Innsbruck University, the Austrian Science Fund (FWF), and the Austrian Federal Ministry for Science, Research and Economy, the University of Adelaide and the Australian Research Council, the Japan Society for the Promotion of Science and by the University of Amsterdam.
We appreciate the excellent work of the technical support staff in Berlin, Durham, Hamburg, Heidelberg, Palaiseau, Paris, Saclay, and in Namibia in the construction and operation of the equipment. This work benefited from services provided by the H.E.S.S. Virtual Organisation, supported by the national resource providers of the EGI Federation.

\end{document}